\documentclass[11pt]{article}
\usepackage{amssymb}

\usepackage{amsmath}

\newcounter{resultnum}[section]

\newcounter{conclusionnum}[section]

\newcounter{conditionnum}[section]

\newcounter{conjecturenum}[section]

\newcounter{examplenum}[section]

\newcounter{exercisenum}[section]
\newtheorem{lemma}{Lemma}[section]

\newcounter{lemmanum}[section]

\newcounter{notationnum}[section]
\newtheorem{theorem}{Theorem}[section]

\newcounter{theoremnum}[section]

\newcounter{definitionnum}[section]
\newtheorem{corollary}{Corollary}[section]

\newcounter{corollarynum}[section]

\newcounter{remarknum}[section]
\newtheorem{proposition}{Proposition}[section]

\newcounter{propositionnum}[section]

\newcounter{acknowledgementnum}[section]

\newcounter{algorithmnum}[section]

\newcounter{axiomnum}[section]

\newcounter{casenum}[section]

\newcounter{claimnum}[section]

\newcounter{summarynum}[section]

\newcounter{problemnum}[section]
\newenvironment{proof}[1][]{\textbf{Proof.} }{}

\begin{document}

\title{Deformation Quantization of Almost K\"{a}hler Models and
Lagrange--Finsler Spaces}
\date{November 15, 2007}
\author{ Sergiu I. Vacaru\thanks{%
sergiu$_{-}$vacaru@yahoo.com, svacaru@fields.utoronto.ca } \\
{\quad} \\
\textsl{The Fields Institute for Research in Mathematical Science,} \\
\textsl{222 College Street, 2d Floor, } \textsl{Toronto \ M5T 3J1, Canada} }
\maketitle

\begin{abstract}
Finsler and Lagrange spaces can be equivalently represented as almost K\"{a}%
hler manifolds endowed with a metric compatible canonical distinguished
connection structure generalizing the Levi Civita connection. The goal of
this paper is to perform a natural Fedosov--type deformation quantization of
such geometries. All constructions are canonically derived for regular
Lagrangians and/or fundamental Finsler functions on tangent bundles.


\vskip3pt


\end{abstract}



\section{ Introduction}

An almost K\"{a}hler manifold (space) is a Riemannian manifold $\mathbb{K}%
=(M^{2n},g)$ of even dimension $2n$ and metric $g$ together with a compatible
almost complex structure $\mathbb{J}$ such that the symplectic form $\theta
\doteqdot g(\mathbb{J}\cdot ,\cdot )$ is closed. Therefore, a space $\mathbb{%
K}$ defines a symplectic geometry with preferred metric and almost complex
structures. In a more general case of almost Hermitian manifolds $\mathbb{H}%
, $ the form $\theta $ is not integrable (not closed), see details in \cite%
{gold,gray}.

Let us consider a smooth $n$--dimensional real manifold $M,$ its tangent
bundle $\mathcal{T}M=\left( TM,\pi ,M\right) $ with surjective projection $%
\pi :TM\rightarrow M,$ total space $TM$ and base $M,$ and denote $\widetilde{%
TM}\doteqdot TM\backslash \{0\},$ where $0$ means the image of the null
cross--section of $\pi .$\footnote{%
For coordinates  $u^{\alpha }=(x^{i},y^{a})$ on $TM,$ when indices
$i,j,..a,..b...$ run values $1,2,...n,$ we get also
coordinates  on $\widetilde{TM}$ if not all fiber coordinates $y^{a}$ vanish;
  in
brief, we shall write $u=(x,y).$} A Finsler space $F^{n}=(M,F)$ consists of
a Finsler metric (fundamental function) $F(x,y)$ defined as a real valued
function $F:TM\rightarrow \mathbb{R}$ with the properties that the
restriction of $F$ to $\widetilde{TM}$ is a function 1) positive ; 2) of
class $C^{\infty }$ and $F$ is only continuous on $\{0\};$ 3) positively
homogeneous of degree 1 with respect to $y^{i},$ i.e. $F(x,\lambda
y)=|\lambda | F(x,y),$\ $\lambda \in \mathbb{R};$
 and 4) the Hessian $\
^{F}g_{ij}=(1/2)\partial ^{2}F/\partial y^{i}\partial y^{j},$ defined on $%
\widetilde{TM},$ is positive definite. There is a classical result by M.
Matsumoto \cite{mats} that a Finsler geometry $F^{n}$ can be modelled as an
almost K\"{a}hler space $^{F}\mathbb{K}=(TM,\ ^{F}\theta ),$ where $\
^{F}\theta \doteqdot \ ^{F}g(\mathbb{J}\cdot ,\cdot )$. for $\mathbb{J}$
adapted to a canonical Finsler nonlinear connection structure.

A Lagrange space $L^{n}=(M,L)$ is defined by a Lagrange fundamental function
$L(x,y)$, i.e. a regular real function $L:$ $TM\rightarrow \mathbb{R},$ for
which the Hessian
\begin{equation}
^{L}g_{ij}=(1/2)\partial ^{2}L/\partial y^{i}\partial y^{j}  \label{hessl}
\end{equation}%
is not degenerate. The concept was introduced by J. Kern \cite{kern} and
developed as a model of geometric mechanics by using methods of Finsler
geometry in a number of works of R. Miron's school on Lagrange--Finsler
geometry and generalizations \cite{ma1987,ma,miat1,miat2,mhl,mhf,bmir}. For
researches interested in applications to modern physics, we note that in our
approach the review \cite{vrfg} is a reference for everything concerning the
geometry of nonholonomic manifolds and generalized Lagrange and Finsler
spaces and applications to standard theories of gravity and gauge fields. It should
be emphasized here that
 a Lagrange space $L^{n}$ is a Finsler space $F^{n}$ if and only if
its fundamental function $L$ is positive and two homogeneous with respect
to variables $y^{i},$ i.e. $L=F^{2}.$ For simplicity, in this paper we shall
work in the bulk with Lagrange spaces, considering the Finsler ones to
consist of a more particular, homogeneous, subclass.

This paper is motivated by the results of V. Oproiu \cite{opr1,opr2,opr} who
proved that any Lagrange space can be realized as an almost K\"{a}hler model
$^{L}\mathbb{K}=(TM,\ ^{L}\theta ),$ where $\ ^{L}\theta \doteqdot \ ^{L}g(%
\mathbb{J}\cdot ,\cdot )$ for $\mathbb{J}$ adapted to a canonical nonlinear
connection structure. This allows us to elaborate a Fedosov quantization
scheme \cite{fedosov1,fedosov2,fedosov,grs} for Lagrange and Finsler spaces.
We shall develop for nonholonomic tangent bundles the approach from
Ref. \cite{karabeg1} in order to prove that any Lagrange and/or Finsler
space can be quantized by geometric deformation methods. The geometric
constructions will be performed for canonical nonlinear and linear
connections in Lagrange and Finsler geometry, see similar results in Ref. %
\cite{esv}, for Lagrange--Fedosov spaces and Fedosov nonholonomic manifolds
provided with almost symplectic connection adapted to the nonlinear
connection structure.

For simplicity, we shall work on tangent bundles even the results have a
natural extension for N--anholonomic manifolds, i.e. manifolds provided with
nonintegrable distributions defining nonlinear connection structures. Such
generalizations of Lagrange--Finsler methods to geometric and (in our cases)
nonholonomic deformations will be used for elaborating certain models of
quantum gravity for lifts of the Einstein gravity on the tangent bundle \cite%
{vqgt} and almost K\"{a}hler models of the Einstein gravity constructed on
nonholonomic semi--Riemannian manifolds \cite{vqgd}.

The work is organized as follows:\ In Section 2 we establish our notations
and remember the basic definitions and results on Lagrange and Finsler
geometry, canonical nonlinear and distinguished connections and metric
structures. We show how such geometric models on tangent bundles can be
reformulated equivalently as almost K\"{a}hler nonholonomic structures.
Section 3 is devoted to basic constructions in nonholonomic deformation and
quantization. We consider star products for symplectic manifolds provided
with nonlinear connection structure, define canonical distinguished
connections adapted to the nonlinear connection and related almost complex
structure. There are defined the Fedosov operators for Lagrange--Finsler
spaces. This allows us, in Section 4, to formulate and sketch the proofs of
Fedosov's theorems for such nonholonomic tangent bundles and provide a
deformation quantization of Lagrange spaces. Finally, we compute the
important coefficient $c_0$ of zero degree of cohomology classes of
quantized Lagrange spaces.

\vskip3pt

\textbf{Conventions:}\ We shall use "left--up" and "left--low" labels like
 $\ ^L \mathbf{D}$ and $%
\ _L \mathbf{g}$ in order to emphasize that the geometric object $\mathbf{g}$
is defined canonically by a regular Lagrange function $L.$ A tensor analysis
in Lagrange--Finsler spaces requires a more sophisticate system of
notations, see details in \cite{vrfg}. Moreover, we use Einstein's
summation convention in local expressions. The system of notations
is a general one used in a series of our works on nonholonomic Einstein
spaces, generalized Ricci--Lagrange flows and nonholonomic deformation quantization. \vskip3pt

\textbf{Acknowledgments: } The author is grateful to Professor Vasile Oproiu
for very important discussions and references on almost K\"{a}hler models of
Lagrange spaces.

\section{Almost K\"{a}hl\-er Lagrange Structures}

In this section, we outline briefly the almost K\"{a}hler model of Lagrange
and Finsler spaces \cite{mats,opr1,opr2,opr,ma1987,ma,vrfg}.

A nonlinear connection (N--connection) $\mathbf{N}$ on a tangent bundle $TM$
can be defined by a Whitney sum (nonholonomic distribution)%
\begin{equation}
TTM=hTM\oplus vTM,  \label{whitney}
\end{equation}%
given locally to by a set of coefficients $N_{i}^{a}(x,y)$ defined with
respect to a coordinate basis $\partial _{\alpha }=\partial /\partial
u^{\alpha }=(\partial _{i}=\partial /\partial x^{i},\partial _{a}=\partial
/\partial y^{a})$ and its dual $du^{\beta }=(dx^{j},dy^{b}).$ In a
particular case, we get linear connections for $N_{i}^{a}=\Gamma
_{i b}^{a}(x)y^{b}.$ The curvature of N--connection is defined as the
Neijenhuis tensor
\begin{equation*}
\Omega _{ij}^{a}=\frac{\partial N_{i}^{a}}{\partial x^{j}}-\frac{\partial
N_{j}^{a}}{\partial x^{i}}+N_{i}^{b}\frac{\partial N_{j}^{a}}{\partial y^{b}}%
-N_{j}^{b}\frac{\partial N_{i}^{a}}{\partial y^{b}}.
\end{equation*}

Let $L(x,y)$ be a regular Lagrangian with nondegenerate $^{L}g_{ij}$ (\ref%
{hessl}) and action integral
\begin{equation*}
S(\tau )=\int\limits_{0}^{1}L(x(\tau ),y(\tau ))d\tau
\end{equation*}%
for $y^{k}(\tau )=dx^{k}(\tau )/d\tau ,$ for $x(\tau )$ parametrizing smooth
curves on a manifold $M$ with $\tau \in \lbrack 0,1]$. We can formulate
 certain very important results
on geometrization of Lagrange mechanics\footnote{%
proofs consist straightforward computations}:

\begin{itemize}
\item The Euler--Lagrange equations $\frac{d}{d\tau }\frac{\partial L}{%
\partial y^{i}}-\frac{\partial L}{\partial x^{i}}=0$ are equivalent to the
''nonlinear geodesic'' (equivalently, semi--spray) equations
\begin{equation*}
\frac{d^{2}x^{k}}{d\tau ^{2}}+2G^{k}(x,y)=0,
\end{equation*}%
where
\begin{equation*}
G^{k}=\frac{1}{4}g^{kj}\left( y^{i}\frac{\partial ^{2}L}{\partial
y^{j}\partial x^{i}}-\frac{\partial L}{\partial x^{j}}\right)
\end{equation*}%
defines the canonical N--connection (for Lagrange spaces) $\ $%
\begin{equation}
\ ^{L}N_{j}^{a}=\frac{\partial G^{a}(x,y)}{\partial y^{j}}.  \label{cncl}
\end{equation}

\item The regular Lagrangian $L(x,y)$ defines a canonical (Sasaki type)
metric structure on $\widetilde{TM},$%
\begin{equation}
\ ^{L}\mathbf{g}=\ ^{L}g_{ij}(x,y)\ e^{i}\otimes e^{j}+\ ^{L}g_{ij}(x,y)\
\mathbf{e}^{i}\otimes \ \mathbf{e}^{j},  \label{slm}
\end{equation}%
where the preferred frame structure (defined linearly by $\ ^{L}N_{j}^{a}$)
is $\mathbf{e}_{\nu }=(\mathbf{e}_{i},e_{a}),$ where
\begin{equation}
\mathbf{e}_{i}=\frac{\partial }{\partial x^{i}}-\ ^{L}N_{i}^{a}(u)\frac{%
\partial }{\partial y^{a}}\mbox{ and
}e_{a}=\frac{\partial }{\partial y^{a}},  \label{dder}
\end{equation}%
and the dual frame (coframe) structure is $\mathbf{e}^{\mu }=(e^{i},\mathbf{e%
}^{a}),$ where
\begin{equation}
e^{i}=dx^{i}\mbox{ and }\mathbf{e}^{a}=dy^{a}+\ ^{L}N_{i}^{a}(u)dx^{i},
\label{ddif}
\end{equation}%
satisfying nontrivial nonholonomy relations
\begin{equation}
\lbrack \mathbf{e}_{\alpha },\mathbf{e}_{\beta }]=\mathbf{e}_{\alpha }%
\mathbf{e}_{\beta }-\mathbf{e}_{\beta }\mathbf{e}_{\alpha }=W_{\alpha \beta
}^{\gamma }\mathbf{e}_{\gamma }  \label{anhrel}
\end{equation}%
with (antisymmetric) nontrivial anholonomy coefficients $W_{ia}^{b}=\partial
_{a}N_{i}^{b}$ and $W_{ji}^{a}=\Omega _{ij}^{a}.$\footnote{%
for simplicity, in this work, we shall omit left labels $L$ in formulas if
that will not result in ambiguities; we shall use boldface indices for
spaces and objects provided or adapted to a N--connection structure.}

\item We get a Riemann--Cartan canonical model $^{RC}L$ on $TM$ of Lagrange
space $L^{n}$ if we choose the canonical metrical distinguished connection $%
\widehat{\mathbf{D}}=(hD,vD)=(\widehat{L}_{\ jk}^{i},\widehat{C}_{jc}^{i})$
(in brief, d--connection, which is a linear connection preserving under
parallelism the splitting (\ref{whitney}))
\begin{equation*}
\widehat{\mathbf{\Gamma }}_{\ j}^{i}=\widehat{\mathbf{\Gamma }}_{\ j\gamma
}^{i}\mathbf{e}^{\gamma }=\widehat{L}_{\ jk}^{i}e^{k}+\widehat{C}_{jc}^{i}%
\mathbf{e}^{c},
\end{equation*}%
for $\widehat{L}_{\ jk}^{i}=\widehat{L}_{\ bk}^{a},\widehat{C}_{jc}^{i}=%
\widehat{C}_{bc}^{a}$ in $\widehat{\mathbf{\Gamma }}_{\ b}^{a}=\widehat{%
\mathbf{\Gamma }}_{\ b\gamma }^{a}\mathbf{e}^{\gamma }=\widehat{L}_{\
bk}^{a}e^{k}+\widehat{C}_{bc}^{a}\mathbf{e}^{c},$ and
\begin{equation}
\widehat{L}_{\ jk}^{i}=\frac{1}{2}g^{ih}(\mathbf{e}_{k}g_{jh}+\mathbf{e}%
_{j}g_{kh}-\mathbf{e}_{h}g_{jk}),\widehat{C}_{\ bc}^{a}=\frac{1}{2}%
g^{ae}(e_{b}g_{ec}+e_{c}g_{eb}-e_{e}g_{bc}),  \label{cdc}
\end{equation}%
which are just the generalized Christoffel indices.\footnote{%
we contract "horizontal" and "vertical" indices following the rule:
$i=1$ is $a=n+1;$ $i=2$ is $a=n+2;$ ... $i=n$ is $a=n+n"$}
We note that $^{RC}L$ contains a
nonholonomically induced torsion structure defined by 2--forms%
\begin{equation}
\mathbf{\Omega }^{i}=\widehat{C}_{\ jc}^{i}e^{i}\wedge \mathbf{e}^{c}%
\mbox{
and }\mathbf{\Omega }^{a}=-\frac{1}{2}\Omega _{ij}^{a}e^{i}\wedge
e^{j}+\left( e_{b}N_{i}^{a}-\widehat{L}_{\ bi}^{a}\right) e^{i}\wedge
\mathbf{e}^{b}  \label{nztors}
\end{equation}%
computed from Cartan's structure equations%
\begin{eqnarray}
de^{i}-e^{k}\wedge \widehat{\mathbf{\Gamma }}_{\ k}^{i} &=&-\mathbf{\Omega }%
^{i},\ d\mathbf{e}^{a}-\mathbf{e}^{b}\wedge \widehat{\mathbf{\Gamma }}_{\
b}^{a}=-\mathbf{\Omega }^{a},  \notag \\
d\widehat{\mathbf{\Gamma }}_{\ j}^{i}-\widehat{\mathbf{\Gamma }}_{\
j}^{k}\wedge \widehat{\mathbf{\Gamma }}_{\ k}^{i} &=&-\mathbf{\Omega }%
_{j}^{i}  \label{seq}
\end{eqnarray}%
in which the curvature 2--form is denoted $\mathbf{\Omega }_{j}^{i},$ see
explicit formulas for coefficients in \cite{ma1987,ma,vrfg} and formula (\ref%
{dcurvtb}).

\item In principle, we can work also with the torsionless Levi
Civita connection $\ ^{L}\nabla ,$ constructed for the same metric (\ref{slm}%
) which with respect to N--adapted bases (\ref{dder}) and (\ref{ddif}) is
given by the same coefficients (\ref{cdc}) but subjected to the condition
 that they must solve the
structure equations (\ref{seq}) with $\ ^{\nabla }\mathbf{\Omega }^{i}=\
^{\nabla }\mathbf{\Omega }^{a}=0$ and $\mathbf{\Omega }_{j}^{i}\neq \
^{\nabla }\mathbf{\Omega }_{j}^{i}.$ This provides a Riemann type
geometrization $^{R}L$ on $TM$ of Lagrange space $L^{n}$. We note that $%
\nabla $ does not preserve under parallelism the N--connection splitting (%
\ref{whitney}), i.e. it is not adapted to the N--connection structure
defined canonically by a regular Lagrangian $L.$ From a formal point of
view, we can work equivalently with both type of connections because $%
\widehat{\mathbf{D}}$ and $^{L}\nabla $ are uniquely defined by the same
data ( $\ ^{L}\mathbf{g}$ and $\ ^{L}N_{j}^{a}),$ i.e. by the same
fundamental function $L,$ in metric compatible forms, $\widehat{\mathbf{D}}%
\mathbf{\ }\ ^{L}\mathbf{g}=hD\ ^{L}\mathbf{g}=vD\ ^{L}\mathbf{g}=0$ and $%
^{L}\nabla \ ^{L}\mathbf{g}=0;$ even $\widehat{\mathbf{D}}$ has a nonzero
torsion, it is induced canonically by the same $\ ^{L}\mathbf{g}$ and $\
^{L}N_{j}^{a}.$
\end{itemize}

The canonical N--connection $\ ^{L}N_{j}^{a}$ (\ref{cncl}) induces an almost K%
\"{a}hler structure defined canonically by a regular $L(x,y)$ \cite%
{opr1,opr2,opr} (in this paper, we use the constructions from \cite%
{ma1987,ma,vrfg}). We introduce an almost complex structure for $L^{n}$ as a
linear operator $\mathbf{J}$ acting on the vectors on $TM$ following
formulas
\begin{equation*}
\mathbf{J}(\mathbf{e}_{i})=-e_{i}\mbox{\ and \ }\mathbf{J}(e_{i})=\mathbf{e}%
_{i},
\end{equation*}%
where the superposition $\mathbf{J\circ J=-I,}$ for $\mathbf{I}$ being the
unity matrix. The operator $\mathbf{J}$ reduces to a complex structure $%
\mathbb{J}$ if and only if the distribution (\ref{whitney}) is integrable.

A regular Lagrangian $L(x,y)$ induces a canonical 1--form
\begin{equation*}
\ ^{L}\omega =\frac{1}{2}\frac{\partial L}{\partial y^{i}}e^{i}
\end{equation*}%
and metric $\ ^{L}\mathbf{g}$ (\ref{slm}) induces a canonical 2--form
\begin{equation}
\ ^{L}\mathbf{\theta }=\ ^{L}g_{ij}(x,y)\mathbf{e}^{i}\wedge e^{j}.
\label{asstr}
\end{equation}%
associated to $\mathbf{J}$ following formulas $\ ^{L}\mathbf{\theta (X,Y)}%
\doteqdot \ ^{L}\mathbf{g}\left( \mathbf{JX,Y}\right) $ for any vectors $%
\mathbf{X}$ and $\mathbf{Y}$ on $TM$ decomposed with respect to a N--adapted
basis (\ref{dder}).

We can prove the results:

\begin{enumerate}
\item A regular $L$ defines on $TM$ an almost Hermitian
(symplectic) structure $\ ^{L}\mathbf{\theta }$ for which
$d\ ^{L}\mathbf{\omega } =\ ^{L}\mathbf{\theta} ;$

\item The canonical N--connection $\ ^{L}N_{j}^{a}$ (\ref{cncl}) and its
curvature have the properties
\begin{equation*}
\sum\limits_{ijk}\ ^{L}g_{l(i}\ ^{L}\Omega _{jk)}^{l}=0,\ \
^{L}g_{ij\shortparallel k}-\ ^{L}g_{ik\shortparallel j}=0,\ e_{k}\
^{L}g_{ij}-e_{j}\ ^{L}g_{ik}=0,
\end{equation*}%
where $(ijk)$ means symmetrization of indices and $$\ ^{L}g_{ij\shortparallel
k}=\mathbf{e}_{k}\ ^{L}g_{ij}-\ ^{L}B_{ik}^{s}\ ^{L}g_{sj}-\ ^{L}B_{jk}^{s}\
^{L}g_{is},$$ for $\ ^{L}B_{ik}^{s}=e_{i}\ ^{L}N_{k}^{s},$ which means that
the almost Hermitian model of a Lagrange space $L^{n}$ is an almost K\"{a}%
hler manifold with $d\ ^{L}\mathbf{\theta }=0.$ We conclude that the triad $%
\mathbb{K}^{2n}=(\widetilde{TM},\ ^{L}\mathbf{g,J})$ defines an almost K\"{a}%
hler space (see details in \cite{opr1,opr2,opr}).
\end{enumerate}

Proofs of properties 1 and 2 follow from computation
\begin{eqnarray*}
d\ ^{L}\mathbf{\theta } &=&\frac{1}{6}\sum\limits_{(ijk)}\ ^{L}g_{is}\
^{L}\Omega _{jk}^{s}e^{i}\wedge e^{j}\wedge e^{k}+\frac{1}{2}\left( \
^{L}g_{ij\parallel k}-\ ^{L}g_{ik\parallel j}\right) \mathbf{e}^{i}\wedge
e^{j}\wedge e^{k} \\
&&+\frac{1}{2}\left( e_{k}\ ^{L}g_{ij}-e_{i}\ ^{L}g_{kj}\right) \mathbf{e}%
^{k}\wedge \mathbf{e}^{i}\wedge e^{j}.
\end{eqnarray*}

The next step is to define the concept of almost K\"{a}hler d--connection $\
^{\theta }\mathbf{D,}$ which is compatible both with the almost K\"{a}hler $%
\left( \ ^{L}\mathbf{\theta ,J}\right) $ and N--connection structures $\ ^{L}%
\mathbf{N,}$ and satisfies the conditions%
\begin{equation*}
\ ^{\theta }\mathbf{D}_{\mathbf{X}}\ ^{L}\mathbf{g=0}\mbox{\ and \ }\mathbf{%
\ }^{\theta }\mathbf{D}_{\mathbf{X}}\mathbf{J=0,}
\end{equation*}%
for any vector $\mathbf{X}=X^{i}\mathbf{e}_{i}+X^{a}e_{a}.$ By a
straightforward computation, we prove (see details in \cite{ma1987,ma}):

\begin{theorem}
\label{th1}The canonical d--connection $\widehat{\mathbf{\Gamma }}_{\ \beta
\gamma }^{\alpha }=\left( \widehat{L}_{\ bk}^{a},\widehat{C}_{bc}^{a}\right)
$\textbf{\ }with coefficients (\ref{cdc}) defines also a (unique) canonical
almost K\"{a}hler d--connection $^{\theta }\widehat{\mathbf{D}}=\widehat{%
\mathbf{D}}$ for which, with respect to N--adapted frames (\ref{dder}) and (%
\ref{ddif}), the coefficients $\widehat{T}_{jk}^{i}=0,$ torsion vanishes on $%
hTM,$ and $\widehat{T}_{bc}^{a}=0,$ torsion vanishes on $vTM,$ but there are
cross non--zero coefficients of type (\ref{nztors}), $%
\widehat{T}_{jc}^{i}=\widehat{C}_{\ jc}^{i},$ $\widehat{T}_{ij}^{a}=\Omega
_{ij}^{a}$ and $\widehat{T}_{ib}^{a}=e_{b}N_{i}^{a}-\widehat{L}_{\ bi}^{a}.$
\end{theorem}

There are two important particular cases: If $L=F^{2},$ for a Finsler space,
we get a almost K\"{a}hler model of Finsler space \cite{mats}, when $%
^{\theta }\widehat{\mathbf{D}}=\widehat{\mathbf{D}}$ transforms in the
so--called Cartan--Finsler connection \cite{cart}. We get a K\"{a}hlerian
model of a Lagrange, or Finsler, space if the respective almost complex
structure $\mathbf{J}$ is integrable.

\section{Nonholonomic Deformations and Quantization}

The geometry of Lagrange--Fedosov manifolds was investigated in Ref. \cite%
{esv}. The aim of this section is to provide a nonholonomic modification of
Fedosov's constructions in order to perform a geometric quantization of
Lagrange (in particular, Finsler) spaces provided with canonical metric and
nonlinear and linear connection structures defined by respective fundamental
Lagrange (Finsler) functions, see next section. We shall use the approach to
Fedosov quantization of geometries with arbitrary metric compatible affine
connections on almost K\"{a}hler manifolds and related symplectic structures
for a manifolds $M$ elaborated in Ref. \cite{karabeg1}. We shall redefine
the constructions from $M$ and $TM,$ respectively, on $TM$ and $TTM$ endowed
with canonical N--connection, metric, symplectic and almost K\"{a}hler
structures uniquely defined by fundamental Lagrange (Finsler) functions.

\subsection{Star products for symplectic manifolds}

Let us denote by $C^{\infty }(V)[[v]]$ the spaces of formal series in
variable $v$ with coefficients from $C^{\infty }(V)$ on a Poisson manifold $%
(V,\{\cdot ,\cdot \}).$ Following Refs. \cite{bffls1,bffls2,bffls3}, a
deformation quantization is an associative algebra structure on $C^{\infty
}(V)[[v]]$ with a $v$--linear and $v$--adically continuous star product
\begin{equation}
\ ^{1}f\ast \ ^{2}f=\sum\limits_{r=0}^{\infty }\ _{r}C(\ ^{1}f,\ ^{2}f)\
v^{r},  \label{starp}
\end{equation}%
where $\ _{r}C,r\geq 0,$ are bilinear operators on $C^{\infty }(V)$ with $\
_{0}C(\ ^{1}f,\ ^{2}f)=\ ^{1}f\ ^{2}f$ and $\ _{1}C(\ ^{1}f,\ ^{2}f)-\
_{1}C(\ ^{2}f,\ ^{1}f)=i\{\ ^{1}f,\ ^{2}f\},$ with $i$ being the complex
unity. Following conventions from \cite{vrfg}, we use ''up'' and ''low''
left labels which are convenient to be introduced on Finsler like spaces in
order to not create confusions with a number of ''horizontal'' and
''vertical'' indices and labels which must be distinguished if the manifolds
are provided with N--connection structure. We note that, in our further
constructions, the manifold $V$ will be a tangent bundle, $V=TM,$ or a
nonhlonomic manifold $V=\mathbf{V,}$ (for instance, a Riemann--Cartan
manifold) provided with a nonholonomic distribution defining a N--connection.

If all operators $\ _{r}C,r\geq 0$ are bidifferential, a corresponding star
product $\ast $ is called differential. We can define different star
products on a $(V,\{\cdot ,\cdot \}).$ Two differential star products $\ast $
and $\ast ^{\prime }$ are equivalent if there is an isomorphism of
algebras $A:\left( C^{\infty }(V)[[v]],\ast \right) \rightarrow \left(
C^{\infty }(V)[[v]],\ast ^{\prime }\right) ,$ where $A=\sum\limits_{r\geq
1}^{\infty }\ _{r}A\ v^{r},$ for $_{0}A$ being the identity operator and $\
_{r}A$ being differential operators on $C^{\infty }(V).$

For a particular case of Poisson manifolds, when $(V,\theta )$ is a
symplectic manifold, each differential star product $\ast $ on $V$ belongs to
its characteristic class
$cl(\ast )\in (1/iv)[\theta ]+H^{2}(V,\mathbb{C})[[v]],$
where $\mathbb{C}$ is the field of complex numbers, and $(1/iv)[\theta
]+H^{2}(V,\mathbb{C})[[v]]$ is an affine vector space, see details in \cite%
{fedosov,deligne,nest}. For symplectic structures, the equivalence classes of
differential star products on $(V,\theta )$ can be bijectively parametrized
by the elements of the corresponding affine vector space using the map $\ast
\rightarrow cl(\ast ).$

The bibliography on existence proofs, methods and descriptions of equivalent
classes for the first examples of star products (Moyal--Weyl and Wick star
products, asymptotic expansions with a numerical parameter, Planck constant,
$\hbar \rightarrow 0,$ Berezin--Toeplitz deformation quantization) is
outlined in Refs. \cite{karabeg1,berez,bms,dls,karabeg2,karabeg3}. A very
important conclusion following from the above--mentioned works
is that  a natural deformation quantization
 can be constructed on an arbitrary compact almost K\"{a}%
hler manifold. The question of existence of deformation quantization and corresponding
geometric formalism on general Poisson manifolds were solved in the
Kontsevich's work \cite{konts1,konts2}. In this work, we are interested to
describe this deformation quantization for Lagrange--Finsler spaces defined
by corresponding nonholonomic structures on arbitrary almost K\"{a}hler
spaces.

\subsection{Canonical d--connections and complex structures}

The deformation quantization using Fedosov's machinery can be also
constructed for a natural class of affine connections, in general,
considered by Yano \cite{yano}. For Lagrange--Finsler spaces and their
almost K\"{a}hler models, we work with the canonical d--connection (\ref{cdc}%
):  the constructions will be re--defined with respect to N--adapted
bases in a form when the results from \cite{karabeg1} will hold true for
nontrivial nonholonomic structures.

Let $\mathbb{K}^{2n}=(\widetilde{TM},\ ^{L}\mathbf{g,J})$ defines an almost K%
\"{a}hler model of Lagrange space with canonical d--connection $^{\theta }%
\widehat{\mathbf{D}}=\widehat{\mathbf{D}}.$ For a chart $U\subset TM,$ we
set the local coordinates $u^{\alpha }=(x^{i},y^{a})$ and parametrize $%
\mathbf{J}(\mathbf{e}_{i})=-e_{i}$ and $\mathbf{J}(e_{i})=\mathbf{e}_{i},$
for $\mathbf{e}_{\alpha }=(\mathbf{e}_{i},e_{a}),$ and denote
\begin{equation}
\mathbf{J}(\mathbf{e}_{\alpha })=\mathbf{J}_{\alpha }^{\ \alpha ^{\prime }}%
\mathbf{e}_{\alpha ^{\prime }},\mbox{ or }\mathbf{J}(\mathbf{e}_{i})=\mathbf{%
J}_{i}^{\ i^{\prime }}\mathbf{e}_{i^{\prime }}\mbox{ and  }\mathbf{J}(e_{a})=%
\mathbf{J}_{a}^{\ a^{\prime }}e_{a^{\prime }}.  \label{aux0}
\end{equation}%
We also write
\begin{equation}
\mathbf{\theta }_{\alpha \beta }=\mathbf{\theta }(\mathbf{e}_{\alpha },%
\mathbf{e}_{\beta }),\mbox{ or }\theta _{ij}=\mathbf{\theta }(\mathbf{e}_{i},%
\mathbf{e}_{j})\mbox{ and  }\theta _{ab}=\mathbf{\theta }(e_{a},e_{b}),
\label{aux1}
\end{equation}%
corresponding to metric (\ref{slm}) with
\begin{equation}
\ ^{L}\mathbf{g}_{\alpha \beta }=\ ^{L}\mathbf{g}(\mathbf{\mathbf{e}}%
_{\alpha }\mathbf{,\mathbf{e}}_{\beta }),\mbox{ or }\ ^{L}g_{ij}=\ ^{L}%
\mathbf{g}(\mathbf{\mathbf{e}}_{i}\mathbf{,\mathbf{e}}_{j})\mbox{ and  }\
^{L}g_{ab}=\ ^{L}\mathbf{g(}e_{a}\mathbf{,}e_{b}\mathbf{).}  \label{aux2}
\end{equation}%
Using the inverse matrices correspondingly to those ones considered above,
we can write%
\begin{equation*}
\mathbf{J}_{\alpha }^{\ \alpha ^{\prime }}=\ ^{L}\mathbf{g}_{\alpha \beta }\
\theta ^{\beta \alpha ^{\prime }}=\ ^{L}\mathbf{g}^{\alpha ^{\prime }\beta
^{\prime }}\theta _{\beta ^{\prime }\alpha }
\end{equation*}%
or in horizontal and vertical (in brief, h- and v-) components,
\begin{equation*}
J_{i}^{\ i^{\prime }}=\ ^{L}g_{ij}\ \theta ^{ji^{\prime }}=\
^{L}g^{i^{\prime }j^{\prime }}\theta _{j^{\prime }i}\mbox{ and  }J_{a}^{\
a^{\prime }}=\ ^{L}g_{ab}\ \theta ^{ba^{\prime }}=\ ^{L}g^{a^{\prime
}b^{\prime }}\theta _{b^{\prime }a}.
\end{equation*}%
We can define $\mathbf{J}_{\ \alpha ^{\prime }}^{\alpha \ }$ as the inverse
to $\mathbf{J}_{\alpha }^{\ \alpha ^{\prime }}.$ In our further
considerations, we shall omit decompositions into h-- and v--components if
it would be possible to write formulas in a more compactified form with Greek
indices not creating ambiguities in distinguishing the nonholonomic
N--connections structure.

The Nijenhuis tensor $\mathbf{\Omega }$ for the complex structure $\mathbf{J}
$ on $TM$ provided with N--connection $\mathbf{N}$ is defined in the form%
\begin{equation*}
\mathbf{\Omega }(\mathbf{X},\mathbf{Y})\doteqdot \left[ \mathbf{JX,JY}\right]
-\mathbf{J}\left[ \mathbf{JX,Y}\right] -\mathbf{J}\left[ \mathbf{X,Y}\right]
-\left[ \mathbf{X,Y}\right]
\end{equation*}%
where $\mathbf{X}$ and $\mathbf{Y}$ are vector fields on $TM.$\footnote{%
We chose a definition of this tensor as in Ref. \cite{ma} which is with
minus sign comparing to the definition used in \cite{karabeg1}; we also use
a different letter for this tensor, like for the N--connection curvature,
because in our case the symbol ''N'' is used for N--connections.}

Let $\mathbf{D=\{\Gamma _{\alpha \beta }^{\gamma }\}}$ $\ $be any metric
compatible d--connection on $TM$ (it is any affine connection preserving the
splitting (\ref{whitney}) and satisfying $\mathbf{D(g)=0}$ for a given
metric structure $\mathbf{g)}.$ A component calculus with respect to
N--adapted bases (\ref{dder}) and (\ref{ddif}), for $\mathbf{\Omega }(%
\mathbf{e}_{\alpha },\mathbf{e}_{\beta })=\mathbf{\Omega }_{\alpha \beta
}^{\gamma }\mathbf{e}_{\gamma }$ results in
\begin{equation}
\mathbf{\Omega }_{\alpha \beta }^{\gamma }=4\mathbf{T}_{\alpha \beta
}^{\gamma }  \label{torsform}
\end{equation}%
where $\mathbf{T}_{\alpha \beta }^{\gamma }$ is the torsion of $\mathbf{%
\Gamma _{\alpha \beta }^{\gamma }.}$\footnote{%
This formula is a nonholonomic analog, for our conventions, with inverse
sign, of the formula (2.9) from \cite{karabeg1}.} For the case of Lagrange
spaces with canonical d--connec\-ti\-on $\widehat{\mathbf{D}}$ (\ref{cdc}),
the non--trivial components $\widehat{\mathbf{T}}_{\alpha \beta }^{\gamma }$
of the are given by nonzero components of 2--forms (\ref{nztors}) being
defined completely by $\ ^{L}g_{ij}$ and $\ ^{L}N_{i}^{a},$ in their turn,
generated by a regular $L(x,y).$

\begin{proposition}
\label{pr1}Any metric compatible d--connection $\mathbf{D}$ with the torsion
given by formula (\ref{torsform}) respects the symplectic form $\mathbf{%
\theta (X,Y)}\doteqdot \mathbf{g}\left( \mathbf{JX,Y}\right) $ and therefore
the complex structure $\mathbf{J.}$
\end{proposition}

\begin{proof}
It consists a straightforward verification of conditions $\mathbf{D}%
_{\mathbf{X}}\mathbf{g}=0$ and $\mathbf{\ D}_{\mathbf{X}}\mathbf{J} =0$
which is a N--adapted calculus with respect to (\ref{dder}) and (\ref{ddif}%
), see similar computations proving the Proposition 2.1 in Ref. \cite%
{karabeg1}. In our case, we work not with affine metric compatible
connections but with respective d--connections. $\Box $
\end{proof}

As a consequence of Theorem \ref{th1} and Proposition \ref{pr1}, we have:

\begin{corollary}
The unique metric compatible canonical d--connection $\widehat{\mathbf{%
\Gamma }}_{\ \beta \gamma }^{\alpha }$ $ = \left( \widehat{L}_{\ bk}^{a},\widehat{%
C}_{bc}^{a}\right) $\textbf{\ }(\ref{cdc}), with torsion components $%
\widehat{T}_{jk}^{i}=0,$ $\widehat{T}_{bc}^{a}=0,$ $\widehat{T}_{jk}^{i}=%
\widehat{C}_{\ jc}^{i},\widehat{T}_{ij}^{a} = \Omega _{ij}^{a}$ and $\widehat{T%
}_{ib}^{a}=e_{b}N_{i}^{a}-\widehat{L}_{\ bi}^{a}$ computed with respect to
N--adapted bases (\ref{dder}) and (\ref{ddif}), satisfies the formula $%
\mathbf{\Omega }_{\alpha \beta }^{\gamma }=4\widehat{\mathbf{T}}_{\alpha
\beta }^{\gamma }$ and respects the canonical symplectic structure $\ ^{L}%
\mathbf{\theta }$ (\ref{asstr}) constructed for the canonical N--connection $%
\ ^{L}N_{j}^{a}$ (\ref{cncl}) and \ metric $\ ^{L}\mathbf{g}$ (\ref{slm}).
In a particular case, for $L=F^{2},$ similar results hold true for the
Cartan connection on Finsler spaces and respective canonical Finsler
symplectic, metric and N--connection structures.
\end{corollary}

We conclude that geometrizing a Lagrange, or Finsler, space in terms of
geometric objects of a nonholonomic almost K\"{a}hler manifold we can
perform directly a natural deformation quantization by adapting the
constructions from Ref. \cite{karabeg1} to the canonical N--connection, and
for respective canonical geometrical objects (like metric, symplectic form,
d--connection, induced by fundamental Lagrange, or Finsler, functions).

Our further considerations will consist from a generalization to the
nonholonomic tangent bundles $TM$ of the results and methods elaborated in
the above--mentioned reference for usual manifolds $M$ endowed with metric
compatible affine connections and respective almost K\"{a}hler structures.
In other turn, we shall emphasize the constructions only for the case of
canonical Lagrange, Finsler, geometric objects. For simplicity, we shall
omit details for local computations and proofs if they will be certain
''N--adapted'' Lagrange--Finsler analogs of formulas and results obtained in %
\cite{karabeg1,yano} or \cite{fedosov1,fedosov2,fedosov,grs}.

\subsection{Fedosov operators for Lagrange--Finsler spaces}

In this section, we modify Fedosov's constructions to provide all data
necessary for deformation quantization of almost K\"{a}hler models of
Lagrange--Finsler spaces.

On $TM$ endowed with canonical Lagrange structures, we introduce the tensor $%
\ ^{L}\mathbf{\Lambda }^{\alpha \beta }\doteqdot \ ^{L}\theta ^{\alpha \beta
}-i\ ^{L}\mathbf{g}^{\alpha \beta },$ see related formulas (\ref{slm}), (\ref%
{asstr}), (\ref{aux1}) and (\ref{aux2}). The local coordinates on $TM$ are
parametrized in the form $u=\{u^{\alpha }\}$ and the local coordinates on $%
T_{u}TM$ are labelled $(u,z)=(u^{\alpha },z^{\beta }),$ where $z^{\beta }$
are the second order fiber coordinates. We shall work with formal series
\begin{equation}
a(v,z)=\sum\limits_{r\geq 0,|\overbrace{\alpha }|\geq 0}\ a_{r,\overbrace{%
\alpha }}(u)z^{\overbrace{\alpha }}\ v^{r},  \label{formser}
\end{equation}%
where $\overbrace{\alpha }$ is a multi--index, defining the formal Wick
algebra $\mathbf{W}_{u}$ (we use a boldface letter in order to emphasize
what we perform our constructions for spaces provided with N--connection
structure). We use the formal Wick product on $\mathbf{W}_{u},$ for two
elements $a$ and $b$ defined by formal series of type (\ref{formser}),%
\begin{equation}
a\circ b\ (z)\doteqdot \exp \left( i\frac{v}{2}\ ^{L}\mathbf{\Lambda }%
^{\alpha \beta }\frac{\partial ^{2}}{\partial z^{\alpha }\partial
z_{[1]}^{\alpha }}\right) a(z)b(z_{[1]})\mid _{z=z_{[1]}}.  \label{fpr}
\end{equation}

It is possible to construct a nonholonomic bundle $\mathbf{W}=\cup _{u}%
\mathbf{W}_{u}$ of formal Wick algebras defined as a union of algebras $%
\mathbf{W}_{u} $ distinguished by the N--connection structure, see Refs. %
\cite{vesnc,vclalg} on such ''d--algebras'', for instance, on gauge and
spinor field geometries adapted to N--connection structures. The fibre
product (\ref{fpr}) can be trivially extended to the space of $\mathbf{W}$%
--valued N--adapted differential forms $\ ^{L}\mathcal{W}\otimes \Lambda $
by means of the usual exterior product of the scalar forms $\mathbf{\Lambda }%
,$ where $\ ^{L}\mathcal{W}$ denotes the sheaf of smooth sections of $%
\mathbf{W}$ (we put the left label $L$ in order to emphasize that the
constructions are adapted to the canonical N--connection structure induced
by a regular Lagrangian). There is a standard grading on $\mathbf{\Lambda ,}$
denoted $\deg _{a}.$ It is possible to introduce grading $\deg _{v},\deg
_{s},\deg _{a}$ on $\ ^{L}\mathcal{W}\otimes \Lambda $ defined on
homogeneous elements $v,z^{\alpha },\mathbf{e}^{\alpha }$ as follows: $\deg
_{v}(v)=1,$ $\deg _{s}(z^{\alpha })=1,$ $\deg _{a}(\mathbf{e}^{\alpha })=1,$
and all other gradings of the elements $v,z^{\alpha },\mathbf{e}^{\alpha }$
are set to zero. In this case, the product $\circ $ from (\ref{fpr}) on $\
^{L}\mathcal{W}\otimes \mathbf{\Lambda }$ is bigraded, we write w.r.t the
grading $Deg=2\deg _{v}+\deg _{s}$ and the grading $\deg _{a},$ see also
conventions from \cite{karabeg1}.

The canonical d--connection $\ ^{L}\widehat{\mathbf{D}}\mathbf{=\{}\ ^{L}%
\widehat{\mathbf{\Gamma }}\mathbf{_{\alpha \beta }^{\gamma }\}}$ can be
extended to an operator on $\ ^{L}\mathcal{W}\otimes \Lambda $ following the
formula
\begin{equation}
\ ^{L}\widehat{\mathbf{D}}\left( a\otimes \lambda \right) \doteqdot \left(
\mathbf{e}_{\alpha }(a)-u^{\beta }\ ^{L}\widehat{\mathbf{\Gamma }}\mathbf{%
_{\alpha \beta }^{\gamma }\ }^{z}\mathbf{e}_{\alpha }(a)\right) \otimes (%
\mathbf{e}^{\alpha }\wedge \lambda )+a\otimes d\lambda ,  \label{cdcop}
\end{equation}%
where $\mathbf{e}_{\alpha }$ and $\mathbf{e}^{\alpha }$ are defined
respectively by formulas (\ref{dder}) and (\ref{ddif}) and $^{z}\mathbf{e}%
_{\alpha }$ is a similar to $\mathbf{e}_{\alpha }$ on N--anholonomic fibers
of $TTM,$ depending on $z$--variables (for holonomic second order fibers, we
can take $^{z}\mathbf{e}_{\alpha }=\partial /\partial z^{\alpha }).$ For
second order mechanical, or Finsler, models, $\mathbf{\ }^{z}\mathbf{e}%
_{\alpha }$ can be constructed canonically from higher order Lagrangians and
respective semi--spray configurations and N--connections \cite%
{miat1,miat2,mhl,mhf,bmir}. In superstring theory and nonholonomic
(super)gravity and higher order spinor structures, such effective higher
order N--connections have to be defined from (super) vielbein configurations %
\cite{vncsup,vhs}. It should be noted that the operator (\ref{cdcop}) can be
similarly defined for arbitrary metric compatible d--connection $\mathbf{%
D=\{\Gamma _{\alpha \beta }^{\gamma }\}}$ and arbitrary N--connection
structures on $TTM,$ but for purposes of this paper we consider only the
case of geometric objects induced canonically by a fundamental function $L,$
or $F.$ Using formulas (\ref{fpr}) and (\ref{cdcop}), we can show that $\
^{L}\widehat{\mathbf{D}}$ is a N--adapted $\deg _{a}$--graded derivation of
the distinguished algebra $\left( \ ^{L}\mathcal{W}\otimes \mathbf{\Lambda
,\circ }\right) ,$ in brief, one call d--algebra.

Now, we can introduce on $\ ^{L}\mathcal{W}\otimes \mathbf{\Lambda }$ the
Fedosov operators $\ ^{L}\delta $ and $\ ^{L}\delta ^{-1}$ (we put
additional left labels in order to emphasize that in this work they are
completely generated by a regular Lagrange, or Finsler, canonical structure
on $TM):$%
\begin{eqnarray*}
\ ^{L}\delta (a) &=&\mathbf{e}^{\alpha }\wedge \mathbf{\ }^{z}\mathbf{e}%
_{\alpha }(a), \\
\ ^{L}\delta ^{-1}(a) &=&\left\{
\begin{array}{c}
\frac{i}{p+q}z^{\alpha }\mathbf{e}_{\alpha }(a),\mbox{ if }p+q>0, \\
{\qquad 0},\mbox{ if }p=q=0,%
\end{array}%
\right.
\end{eqnarray*}%
where $a\in \ ^{L}\mathcal{W}\otimes \mathbf{\Lambda }$ is homogeneous
w.r.t. the grading $\deg _{s}$ and $\deg _{a}$ with $\deg _{s}(a)=p$ and $%
\deg _{a}(a)=q.$ We get the formula
\begin{equation*}
a=(\ ^{L}\delta \ ^{L}\delta ^{-1}+\ ^{L}\delta ^{-1}\ ^{L}\delta +\sigma
)(a)
\end{equation*}%
where $a\longmapsto \sigma (a)$ is the projection on the $(\deg _{s},\deg
_{a})$--bihomogeneous part of $a$ of degree zero, $\deg _{s}(a)=\deg
_{a}(a)=0.$ We can verify that $^{L}\delta $ is also a $\deg _{a}$%
--graded derivation of the d--algebra $\left( \ ^{L}\mathcal{W}\otimes
\mathbf{\Lambda ,\circ }\right) .$

The Lagrange canonical d--connection $^{L}\widehat{\mathbf{D}}$ (\ref{cdc})
on $TM$ induces respective torsion and curvature on $\ ^{L}\mathcal{W}%
\otimes \mathbf{\Lambda ,}$%
\begin{equation}
\widehat{\mathcal{T}}\doteqdot \frac{z^{\gamma }}{2}\ ^{L}\theta _{\gamma
\tau }\widehat{\mathbf{T}}_{\alpha \beta }^{\tau }(u)\mathbf{e}^{\alpha
}\wedge \mathbf{e}^{\beta }  \label{at1}
\end{equation}%
and
\begin{equation}
\widehat{\mathcal{R}}\doteqdot \frac{z^{\gamma }z^{\varphi }}{4}\ ^{L}\theta
_{\gamma \tau }\widehat{\mathbf{R}}_{\ \varphi \alpha \beta }^{\tau }(u)%
\mathbf{e}^{\alpha }\wedge \mathbf{e}^{\beta }  \label{ac1}
\end{equation}%
where the torsion $\widehat{\mathbf{T}}_{\alpha \beta }^{\tau }$ (\ref%
{nztors}) has nontrivial components $\widehat{T}_{jk}^{i}=\widehat{C}_{\
jc}^{i},\widehat{T}_{ij}^{a}=\Omega _{ij}^{a}$ and $\widehat{T}%
_{ib}^{a}=e_{b}N_{i}^{a}-\widehat{L}_{\ bi}^{a}$ and the curvature $\widehat{%
\mathbf{R}}_{\ \varphi \alpha \beta }^{\tau }$ with nontrivial components
\begin{eqnarray}
\widehat{R}_{\ hjk}^{i} &=&\mathbf{e}_{k}\widehat{L}_{\ hj}^{i}-\mathbf{e}%
_{j}\widehat{L}_{\ hk}^{i}+\widehat{L}_{\ hj}^{m}\widehat{L}_{\ mk}^{i}-%
\widehat{L}_{\ hk}^{m}\widehat{L}_{\ mj}^{i}-\widehat{C}_{\ ha}^{i}\Omega
_{\ kj}^{a},  \label{dcurvtb} \\
\widehat{P}_{\ jka}^{i} &=&e_{a}\widehat{L}_{\ jk}^{i}-\widehat{\mathbf{D}}%
_{k}\widehat{C}_{\ ja}^{i},\ \widehat{S}_{\ bcd}^{a}=e_{d}\widehat{C}_{\
bc}^{a}-e_{c}\widehat{C}_{\ bd}^{a}+\widehat{C}_{\ bc}^{e}\widehat{C}_{\
ed}^{a}-\widehat{C}_{\ bd}^{e}\widehat{C}_{\ ec}^{a},  \notag
\end{eqnarray}%
all computed in a forme when there are solved
 the structure equations (\ref{seq}).

Using the formulas (\ref{formser}) and (\ref{fpr}) and the identity
\begin{equation}
\ ^{L}\theta _{\varphi \tau }\widehat{\mathbf{R}}_{\ \gamma \alpha \beta
}^{\tau }=\ ^{L}\theta _{\gamma \tau }\widehat{\mathbf{R}}_{\ \varphi \alpha
\beta }^{\tau },  \label{acp}
\end{equation}%
we prove the formulas%
\begin{equation}
\left[ \ ^{L}\widehat{\mathbf{D}},\ ^{L}\delta \right] =\frac{i}{v}ad_{Wick}(%
\widehat{\mathcal{T}})\mbox{ and }\ ^{L}\widehat{\mathbf{D}}^{2}=-\frac{i}{v}%
ad_{Wick}(\widehat{\mathcal{R}}),  \label{comf}
\end{equation}%
where $[\cdot ,\cdot ]$ is the $\deg _{a}$--graded commutator of
endomorphisms of $\ ^{L}\mathcal{W}\otimes \mathbf{\Lambda }$ and $ad_{Wick}$
is defined via the $\deg _{a}$--graded commutator in $\left( \ ^{L}\mathcal{W%
}\otimes \mathbf{\Lambda ,\circ }\right) .$\footnote{%
It should be noted that formulas (\ref{at1}), (\ref{ac1}), (\ref{dcurvtb}), (%
\ref{acp}) and (\ref{comf}) can be written for any metric\textbf{\ }$\mathbf{%
g}$ and metric compatible d--connection $\mathbf{D,Dg=0,}$ on $TM,$ provided
with arbitrary N--connection \ $\mathbf{N}$ (we have to omit ''hats'' and
labels $L$). It is a more sophisticate problem to define such constructions
for Finsler geometries with the so--called Chern connection which are metric
noncompatible \cite{bcs}. For applications in standard models of physics, we
chose the variants of Lagrange--Finsler spaces defined by metric compatible
d--connections, see discussion in \cite{vrfg}.}

\section{Fedosov Quantization of Lagrange Spaces}

We generalize the standard statements of Fedosov's theory for the case of
Lagrange--Finsler spaces provided with canonical N--connection and
d--con\-nec\-ti\-on structures. The class $c_{0}$ of the deformation
quantization of Lagrange geometry is calculated.

\subsection{Fedosov's theorems for Lagrange--Finsler spaces}

Using the formalism of Fedosov operators on Lagrange spaces, we formulate
and sketch the proof of two theorems generalizing similar ones to the case
of N--connection and metric compatible d--connection geometries on $TM,$
induced by fundamental Lagrange (Finsler) functions.

Let us denote the $Deg$--homogeneous component of degree $k$ of an element $%
a\in $ $\ ^{L}\mathcal{W}\otimes \mathbf{\Lambda }$ by $a^{(k)}.$

\begin{theorem}
\label{th2}For any regular Lagrangian $L$ on $\widetilde{TM},$ there is a
flat canonical Fedosov d--connection
\begin{equation*}
\ ^{L}\widehat{\mathcal{D}}\doteqdot -\ ^{L}\delta +\ ^{L}\widehat{\mathbf{D}%
}-\frac{i}{v}ad_{Wick}(r)
\end{equation*}%
satisfying the condition $\ ^{L}\widehat{\mathcal{D}}^{2}=0,$ where the
unique element $r\in $ $\ ^{L}\mathcal{W}\otimes \mathbf{\Lambda ,}$ $\deg
_{a}(r)=1,$ $\ ^{L}\delta ^{-1}r=0,$ solves the equation
\begin{equation*}
\ ^{L}\delta r=\widehat{\mathcal{T}}+\widehat{\mathcal{R}}+\ ^{L}\widehat{%
\mathbf{D}}r-\frac{i}{v}r\circ r
\end{equation*}%
and this element can be computed recursively with respect to the total
degree $Deg$ as follows:%
\begin{eqnarray*}
r^{(0)} &=&r^{(1)}=0,\ r^{(2)} =\ ^{L}\delta ^{-1}\widehat{\mathcal{T}}, \\
r^{(3)} &=&\ ^{L}\delta ^{-1}\left( \widehat{\mathcal{R}}+\ ^{L}\widehat{%
\mathbf{D}}r^{(2)}-\frac{i}{v}r^{(2)}\circ r^{(2)}\right) , \\
r^{(k+3)} &=&\ ^{L}\delta ^{-1}\left( \ ^{L}\widehat{\mathbf{D}}r^{(k+2)}-%
\frac{i}{v}\sum\limits_{l=0}^{k}r^{(l+2)}\circ r^{(l+2)}\right) ,k\geq 1.
\end{eqnarray*}
\end{theorem}

\begin{proof}
We sketch the idea of proof which is similar to the standard Fedosov
constructions but N--adapted. By induction, we use the identities%
\begin{equation*}
\ ^{L}\delta \widehat{\mathcal{T}}=0\mbox{ and }\ ^{L}\delta \widehat{%
\mathcal{R}}=\ ^{L}\widehat{\mathbf{D}}\widehat{\mathcal{T}}.
\end{equation*}%
In Ref. \cite{karabeg1}, these identities were proved for arbitrary affine
connections with torsion and almost K\"{a}hler structures on $M.$ In our
case, we work with more a particular class of geometric objects, induced
canonically from $L^{n},$ on $TM.$ In another turn, the constructions are
generalized to nonholonomic bundles. $\Box $
\end{proof}

\vskip3pt

We note that the canonical Fedosov d--connection $\ ^{L}\widehat{\mathcal{D}}
$ is a $\deg _{a}$--graded derivation of the algebra $\left( \ ^{L}\mathcal{W%
}\otimes \mathbf{\Lambda ,\circ }\right) .$ This means that $\ ^{L}\mathcal{W%
}_{\widehat{\mathcal{D}}}\doteqdot \ker \left( \ ^{L}\widehat{\mathcal{D}}%
\right) $ $\cap \ ^{L}\mathcal{W}$ is N--adapted subalgebra of $\left( \ ^{L}%
\mathcal{W}\mathbf{,\circ }\right) .$

The next theorem gives a rule how to define and compute the star product
induced by a regular Lagrangian.

\begin{theorem}
\label{th3}A star--product on the canonical almost K\"{a}hler model of
Lagrange (Finsler) spaces $\mathbb{K}^{2n}=(\widetilde{TM},\ ^{L}\mathbf{g,J}%
)$ is defined on $C^{\infty }(\widetilde{TM})[[v]]$ by formula
\begin{equation*}
\ ^{1}f\ast \ ^{2}f\doteqdot \sigma (\tau (\ ^{1}f))\circ \sigma (\tau (\
^{2}f)),
\end{equation*}%
where the projection $\sigma :\ ^{L}\mathcal{W}_{\widehat{\mathcal{D}}%
}\rightarrow C^{\infty }(\widetilde{TM})[[v]]$ onto the part of $\deg _{s}$%
--degree zero is a bijection and the inverse map $\tau :C^{\infty }(%
\widetilde{TM})[[v]]\rightarrow \ ^{L}\mathcal{W}_{\widehat{\mathcal{D}}}$
can be calculated recursively w.r.t. the total degree $Deg,$%
\begin{eqnarray*}
\tau (f)^{(0)} &=&f \mbox{\ and, for } k\geq 0, \\
\tau (f)^{(k+1)} &=&\ ^{L}\delta ^{-1}\left(\ ^{L}\widehat{\mathbf{D}}\tau
(f)^{(k)}-\frac{i}{v}\sum\limits_{l=0}^{k}ad_{Wick}(r^{(l+2)})(\tau
(f)^{(k-l)})\right).
\end{eqnarray*}
\end{theorem}

\begin{proof}
We note that the connection $\ ^{L}\widehat{\mathbf{D}}$ and its almost K%
\"{a}hler version defined by Proposition \ref{pr1}, Theorems \ref{th1} and %
\ref{th2} in the case of almost K\"{a}hler manifolds is a special N--adapted
case of the star--product constructed in Ref. \cite{bord}. $\Box $
\end{proof}

\vskip5pt

The statements of the above presented Fedosov's theorems generalized for
Lagrange--Finsler spaces can be extended for arbitrary metric compatible
d--connections on $\widetilde{TM}.$ For Finsler spaces, we can use the
so--called R. Miron's procedure of computing all metric compatible
d--connections for a given metric $\mathbf{g},$ see Refs. \cite{ma1987,ma}
(from a formal point of view, we shall have the same formulas without ''hats''
and $L$--labels, but with arbitrary d--torsions and corresponding
curvatures). It should be noted that there is also the so--called Kawaguchi
metrization procedure, which allows to work with metric noncompatible
d--connections, described in details in the above--mentioned Miron and
Anastasiei
monographs. In Ref. \cite{vsgg}, such constructions were elaborated for
nonholonomic manifolds with the aim to apply Finsler methods in modern
gravity theories.

\subsection{Cohomology classes of quantized Lagrange spaces}

It follows from the results obtained in \cite{karabeg3} that the
characteristic class of the star product from \cite{bord} is $(1/iv)[\theta
]-(1/2i)\varepsilon ,$ where $\varepsilon $ is the canonical class for an
underlying K\"{a}hler manifold. This canonical class can be defined for any
almost complex manifold. In this section, we calculate the crucial part of
the characteristic class $cl$ of the star product $\ast $ which we have
constructed in Theorem \ref{th3} for an almost K\"{a}hler model of Lagrange
(Finsler) spaces, i.e. we shall compute the coefficient $c_{0}$ at the
zeroth degree of $v.$ Only the coefficient $c_{0}(\ast )$ of the class $%
cl(\ast )=(1/iv)[\ ^{L}\theta ]+c_{0}(\ast )+...$ can not recovered from
Deligne's intrinsic class \cite{deligne}. Here we also note that the
cohomology class of the formal K\"{a}hler form parametrizing a quantization
with separation of variables on a K\"{a}hler manifold differs from the
characteristic class of this quantization only in the coefficient $c_{0}$ as
it is proved in \cite{karabeg3}.

Let us recall the rigorous definition of the class $c_{0}$ of a
star--product (\ref{starp}) (see details, for instance, in Ref. \cite{grs})
adapting the constructions for tangent bundles provided with N--connection
structure. One denotes by $\ ^{f}\xi $ the corresponding Hamiltonian vector
field corresponding to a function $f\in C^{\infty }(TM)$ on a symplectic
tangent bundle $(TM,\theta )$ and considers the antisymmetric part%
\begin{equation*}
\ ^{-}C(\ ^{1}f,\ ^{2}f)\ \doteqdot \frac{1}{2}\left( C(\ ^{1}f,\ ^{2}f)-C(\
^{2}f,\ ^{1}f)\right)
\end{equation*}%
of bilinear operator $C(\ ^{1}f,\ ^{2}f).$ A star--product (\ref{starp}) is
normalized if
\begin{equation*}
\ _{1}C(\ ^{1}f,\ ^{2}f)=\frac{i}{2}\{\ ^{1}f,\ ^{2}f\},
\end{equation*}%
where $\{\cdot ,\cdot \}$ is the Poisson bracket. For normalized $\ast $ the
bilinear operator $\ _{2}^{-}C$ is a de Rham--Chevalley 2--cocycle. In this
case, there is a unique closed 2--form $\ ^{L}\varkappa ,$ induced
 by a regular Lagrangian $L,$ such that%
\begin{equation}
\ _{2}C(\ ^{1}f,\ ^{2}f)=\frac{1}{2}\ ^{L}\varkappa (\ ^{f_{1}}\xi ,\
^{f_{2}}\xi )  \label{c2}
\end{equation}%
for all $\ ^{1}f,\ ^{2}f\in C^{\infty }(TM).$ The class $c_{0}$ of a
normalized star--product $\ast $ is stated as the equivalence class $%
c_{0}(\ast )\doteqdot \lbrack \ ^{L}\varkappa ].$

A fiberwise equivalence operator on $\ ^{L}\mathcal{W}$ can be defined by
the formula
\begin{equation*}
\ ^{L}G\doteqdot \exp \left( -v\ ^{L}\bigtriangleup \right) ,
\end{equation*}%
where
\begin{equation*}
\ ^{L}\bigtriangleup =\frac{1}{8}\ ^{L}\mathbf{g}^{\alpha \beta }\left(
\mathbf{\ }^{z}\mathbf{e}_{\alpha }\mathbf{\ }^{z}\mathbf{e}_{\beta }+%
\mathbf{\ }^{z}\mathbf{e}_{\beta }\mathbf{\ }^{z}\mathbf{e}_{\alpha }\right)
\end{equation*}%
for nonholonomic configurations on the second order fibers on $TTM,$ or
\begin{equation*}
\ ^{L}\bigtriangleup =\frac{1}{4}\ ^{L}\mathbf{g}^{\alpha \beta }\frac{%
\partial ^{2}}{\partial z^{\alpha }\partial z^{\beta }}
\end{equation*}%
if we elaborate a model with trivial N--connection for the second order
fibres on $TTM.$ We can check directly the formulas
\begin{equation}
\left[ \ ^{L}\widehat{\mathbf{D}},\ ^{L}\bigtriangleup \right] =\left[ \ ^{L}%
\widehat{\mathbf{D}},\ ^{L}G\right] =0\mbox{ and }\left[ \ ^{L}\mathbf{%
\delta },\ ^{L}\bigtriangleup \right] =\left[ \ ^{L}\mathbf{\delta },\ ^{L}G%
\right] =0,  \label{aux4}
\end{equation}%
which allows us to define a fibrewise star--product on $\ ^{L}\mathcal{W},$
\begin{equation*}
a\circ ^{\prime }b\doteqdot \ ^{L}G(\ ^{L}G^{-1}a\circ \ ^{L}G^{-1}b),
\end{equation*}%
which is the Weyl star--product%
\begin{eqnarray*}
a\circ ^{\prime }b(z) &=&\exp \left( \frac{iv}{4}\ ^{L}\theta ^{\alpha \beta
}(\mathbf{e}_{\alpha }\mathbf{\ }^{z}\mathbf{e}_{\beta }-\mathbf{\ }^{z}%
\mathbf{e}_{\beta }\mathbf{e}_{\alpha })\right) ; \\
&=&\exp \left( \frac{iv}{2}\ ^{L}\theta ^{\alpha \beta }\left( \mathbf{e}%
_{\alpha }\frac{\partial }{\partial z^{\beta }}-\frac{\partial }{\partial
z^{\beta }}\mathbf{e}_{\alpha }\right) \right) , \\
 && \mbox{for holonomic 2d order fibres}.
\end{eqnarray*}%
The next step is to push forward the Fedosov d--connection $\ ^{L}\widehat{%
\mathcal{D}}$ from Theorem \ref{th2} using formulas (\ref{aux4}). We get a
new canonical d--connection operator%
\begin{equation*}
\ ^{L}\widehat{\mathcal{D}}^{\prime }=\ ^{L}G\ ^{L}\widehat{\mathcal{D}}\
^{L}G^{-1}=\ ^{L}\mathbf{\delta +}\ ^{L}\widehat{\mathbf{D}}-\frac{i}{v}%
ad_{Weyl}(r^{\prime }),
\end{equation*}%
where $r^{\prime }=\ ^{L}Gr$ and $ad_{Weyl}$ is calculated with respect to
the $\circ ^{\prime }$--commutator.

For symplectic manifolds, it is well known that each star--product is
equivalent to a normalized one. The class $c_{0}(\ast )$ of a star product $%
\ast $ is defined as the cohomology class $c_{0}(\ast ^{\prime })$ of an
equivalent normalized star--product $\ast ^{\prime }.$ We have first to
construct an equivalent normalized star--product in order to calculate the
class $c_{0}(\ast )$ for $\ast $ from Theorem \ref{th3}. This procedure is
described in details in section 4 of Ref. \cite{karabeg1} for arbitrary
affine metric compatible connection on a manifold $M.$ In our case, those
formulas have to be redefined with respect to N--adapted bases and canonical
d--connection, N--connection and metric structures. For simplicity, we omit
in this work such tedious but trivial generalizations but present only the
most important formulas and definitions.

A straightforward computation of $\ _{2}C$ from (\ref{c2}), using statements
of Theorem \ref{th2}, results in a proof of

\begin{lemma}
\label{lem1}The unique 2--form $\ ^{L}\varkappa $ can be expressed in the
form
\begin{equation*}
\ ^{L}\varkappa =-\frac{i}{8}\mathbf{J}_{\tau }^{\ \alpha ^{\prime }}%
\widehat{\mathbf{R}}_{\ \alpha ^{\prime }\alpha \beta }^{\tau }\mathbf{e}%
^{\alpha }\wedge \mathbf{e}^{\beta }-i\ ^{L}\lambda ,
\end{equation*}%
where the exact N--adapted 1--form $\ ^{L}\lambda =d\ ^{L}\mu ,$ for
\begin{equation*}
\ ^{L}\mu =\frac{1}{6}\mathbf{J}_{\tau }^{\ \alpha ^{\prime }}\widehat{%
\mathbf{T}}_{\ \alpha ^{\prime }\beta }^{\tau }\mathbf{e}^{\beta },
\end{equation*}%
with nontrivial components of curvature and torsion defined by the canonical
d--connection computed following formulas (\ref{nztors}) and (\ref{dcurvtb}).
\end{lemma}

For trivial N--connection structures and arbitrary metric and metric
compatible affine connections, the Lemma \ref{lem1} is equivalent to
statements of Lemma 4.1 from \cite{karabeg1}, in our case, redefined for Riemann--Cartan
geometries modelled on $TM.$ We reformulated the results in a form when
generalizations for arbitrary metric $\mathbf{g}$ compatible d--connection
and N--connection structures, $\mathbf{D}$ and $\mathbf{N}$ on $TM,$ can be
performed following the formal rule of omitting ''hats'' and ''$L$--labels''.

Let us recall the definition of the canonical class $\varepsilon $ of an
almost complex manifold $(M,\mathbb{J})$ and redefine it for $\
^{N}TTM=hTM\oplus vTM$ (\ref{whitney}) stating a N--connection structure $%
\mathbf{N}.$ The distinguished complexification of such second order tangent
bundles is introduced in the form
\begin{equation*}
T_{\mathbb{C}}\left( \ ^{N}TTM\right) =T_{\mathbb{C}}\left( hTM\right)
\oplus T_{\mathbb{C}}\left( vTM\right) .
\end{equation*}%
For such nonholonomic bundles, the class $\ ^{N}\varepsilon $ is the first
Chern class of the distributions $T_{\mathbb{C}}^{\prime }\left( \
^{N}TTM\right) =T_{\mathbb{C}}^{\prime }\left( hTM\right) \oplus T_{\mathbb{C%
}}^{\prime }\left( vTM\right) $ of couples of vectors of type $(1,0)$ both
for the h-- and v--parts. Our aim is to calculate the canonical class $%
^{L}\varepsilon $ (we put the label $L$ for the constructions canonically
defined by a regular Lagrangian $L)$ for the almost K\"{a}hler model of a
Lagrange space $L^{n}.$ We take the canonical d--connection $\ ^{L}\widehat{%
\mathbf{D}}$ that it was used for constructing $\ast $ and considers h- and
v--projections
\begin{equation*}
h\Pi =\frac{1}{2}(Id_{h}-iJ_{h})\mbox{ and }v\Pi =\frac{1}{2}(Id_{v}-iJ_{v}),
\end{equation*}%
where $Id_{h}$ and $Id_{v}$ are respective identity operators and $J_{h}$
and $J_{v}$ are defined by formulas (\ref{aux0}), which are projection
operators onto corresponding $(1,0)$--subspaces. It follows from (\ref{acp})
that $Tr\widehat{\mathbf{R}}=Tr(\Omega _{\beta }^{\alpha })=0,$ see (\ref%
{seq}). The matrix $\left( h\Pi ,v\Pi \right) \widehat{\mathbf{R}}\left(
h\Pi ,v\Pi \right) ^{T},$ where $(...)^{T}$ means transposition, is the
curvature matrix of the N--adapted restriction of the connection $\ ^{L}%
\widehat{\mathbf{D}}$ to $T_{\mathbb{C}}^{\prime }\left(\ ^{N}TTM\right).$
  Now, we can compute the Chern--Weyl form
\begin{eqnarray*}
\ ^{L}\gamma &=&-iTr\left[ \left( h\Pi ,v\Pi \right) \widehat{\mathbf{R}}%
\left( h\Pi ,v\Pi \right) ^{T}\right] =-iTr\left[ \left( h\Pi ,v\Pi \right)
\widehat{\mathbf{R}}\right] \\
&=&-\frac{1}{4}\mathbf{J}_{\tau }^{\ \alpha ^{\prime }}\widehat{\mathbf{R}}%
_{\ \alpha ^{\prime }\alpha \beta }^{\tau }\mathbf{e}^{\alpha }\wedge
\mathbf{e}^{\beta }
\end{eqnarray*}%
to be closed. By definition, the canonical class is $^{L}\varepsilon
\doteqdot \lbrack \ ^{L}\gamma ].$ It follows from Lemma \ref{lem1} and the
above presented considerations the proof of

\begin{theorem}
The zero--degree cohomology coefficient $c_{0}(\ast )$ for the almost K\"{a}%
hler model of Lagrange space is computed
\begin{equation*}
c_{0}(\ast )=-(1/2i)\ ^{L}\varepsilon ,
\end{equation*}
where the value $\ ^{L}\varepsilon $ is canonically defined by a regular
Lagrangian $L(u).$
\end{theorem}

Finally we note that the formula from this Theorem can be directly applied
for the Cartan connection in Finsler geometry with $L=F^{2}.$ In our partner
works \cite{vqgt,vqgd}, we consider its extensions respectively for
generalized Lagrange spaces, or canonical nonholonomic lifts of
semi--Riemanian metrics on $TM,$ and nonholonomic deformations of the
Einstein gravity into almost K\"{a}hler models on nonholonomic manifolds.

\end{document}